\documentclass{optica-article}
\journal{opticajournal} 
\articletype{Research Article}
\usepackage{lineno}
\usepackage{pifont}

\usepackage{graphicx}
\usepackage{subcaption}
\usepackage{xcolor}

\usepackage{hyperref}
\hypersetup{
    colorlinks=true,
    linkcolor=blue,
    filecolor=blue,
    urlcolor=blue,
   citecolor=blue,
}

\begin{document}

\title{Guided polariton condensate in perovskite microwires}

\author{Maciej Nytko,\authormark{1} Mateusz Kędziora,\authormark{1} Maciej Zaremba,\authormark{1} Marek Ekielski,\authormark{2} Anna Szerling,\authormark{2}  Krzysztof Tyszka,\authormark{1} Barbara Piętka\authormark{1,*} }


\address{\authormark{1}Institute of Experimental Physics, Faculty of Physics, University of Warsaw, Warsaw, Poland\\
\authormark{2}Łukasiewicz Research Network - Institute of Microelectronics and Photonics, Warsaw, Poland}

\email{\authormark{*}barbara.pietka@fuw.edu.pl}

\begin{abstract*}
Perovskite microwires are promising candidates for integrated photonic systems due to their strong nonlinear optical response and inherent waveguiding capabilities. 
In this study, we focus on the directional emission properties of exciton-polariton condensates formed within perovskite microwires, with emphasis on emission collected from the microwire end.
We constructed a multi-angle optical detection setup that allows us to identify the exciton-polariton condensation threshold and track the evolution of the condensate for different excitation position along and across the microwire. 
We observe spectrally narrow exciton-polariton condensate emission from the microwire end even when the condensate is generated tens of micrometers away, which demonstrates the ability of the exciton-polariton condensate to propagate over long distances within the microwire.
Furthermore, we find that the presence of structural defects near the condensate location can significantly enhance the emission from the microwire end due to aligning the condensate's momentum with the waveguide direction, thereby facilitating more efficient propagation.     
\end{abstract*}

\section{Introduction}

Lead-halide perovskite materials have emerged as promising candidates for the development of integrated photonic systems due to their optical properties \cite{PL-efficiency, room-temp, Opala-Matuszewski-ploaritoncomputing, Kumar2022-zastosowania, Sanvitto2016-devices}.
In particular, these materials can support the formation of exciton-polariton condensates \cite{Hopfield1958,Kasprzak2006,PhysRevLett.99.140402} (further referred to as condensates) even at room temperature \cite{room-temp,room-temp2,Kędziora-Nature}.
Those condensates exhibit strong optical nonlinearities \cite{Kędziora-Nature,Su2020-nonlinearroomtemp}, one of the essential features for the implementation of neuromorphic computing architectures \cite{Opala-Matuszewski-ploaritoncomputing,Feng2021}.
Furthermore, the high refractive index of perovskites enables efficient light confinement within the structure, allowing them to function as optical waveguides without the need for additional structuring or cladding beyond direct crystallization \cite{Kędziora-Nature,Zhang2023-pero_waveguides,Liu2022-pero_waveguide,C5NR06262D-waveguide,Li2021-waveguide}.
This dual functionality, serving both as a nonlinear light source and a waveguiding medium, offers the potential to significantly simplify the design and fabrication of compact integrated photonic circuits, eliminating the need to combine multiple materials and processing steps \cite{chrostowski2015silicon,Capmany_Perez2020}.
Thanks to advancing perovskite microwire fabrication methods, we are capable of designing and manufacturing various predefined geometries, such as microwires, couplers, and splitters \cite{Kędziora-Nature}.

In this work we investigate the optical properties of CsPbBr$_3$ perovskite microwires, which serve both as the material in which exciton-polariton condensates are generated and as optical waveguides facilitating their propagation.
By analyzing the efficiency 
of condensate transport along these microwires, 
we uncover key mechanisms that govern long-range polariton flow \cite{Su2018} in perovskite structures.
These insights are essential for advancing the development of compact, all-optical circuits and neuromorphic photonic networks, where controlled condensate transport plays a critical functional role \cite{Sedov2025,Topfer:21}.

\section{Methods}

\begin{figure}[htbp]
\includegraphics[width=0.99\linewidth]{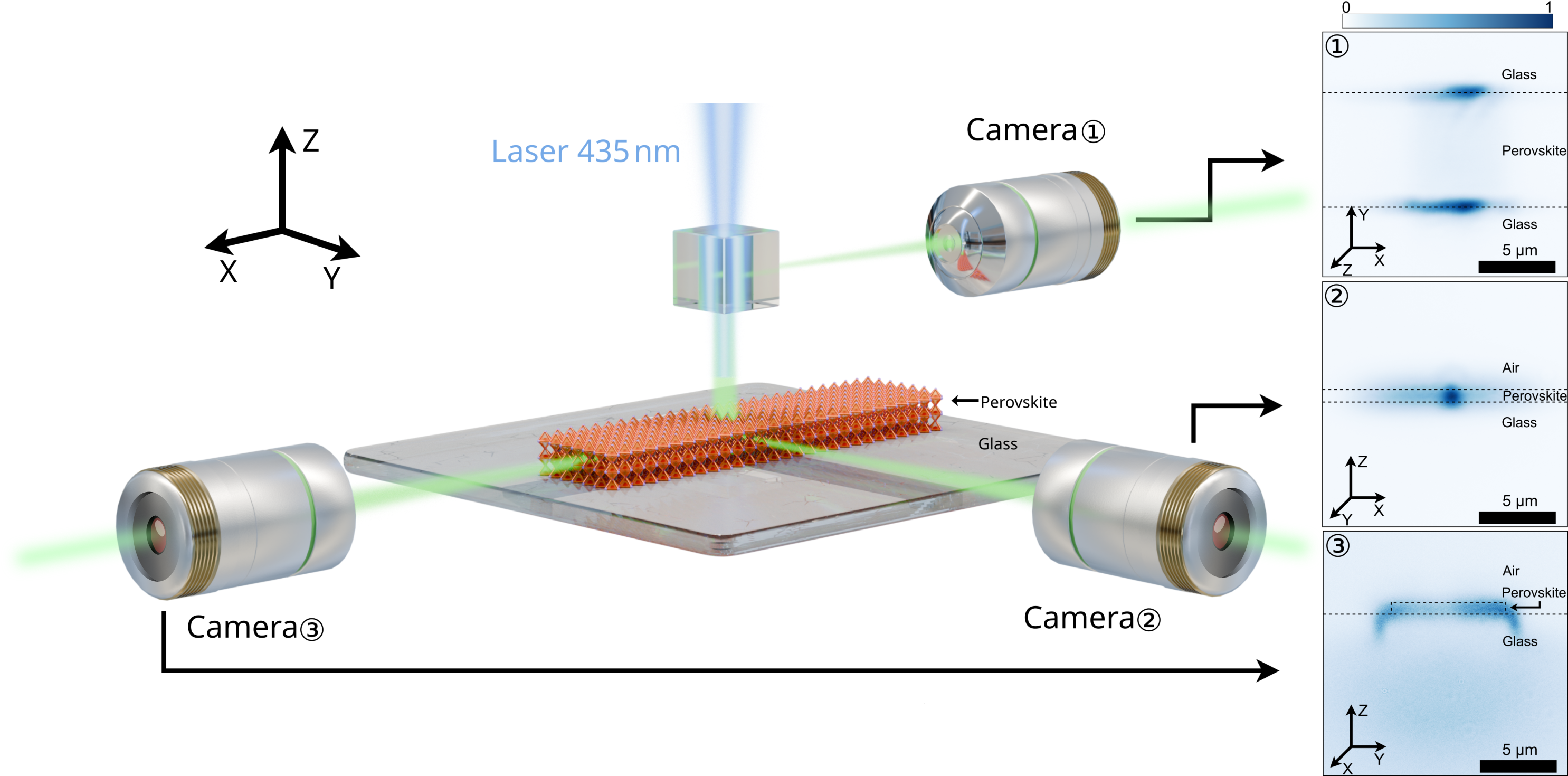}
\caption{Emission detection setup. Perovskite microwire is pumped with 435 nm pulsed laser, and its emission from each side of the perovskite microwire is collected via CCD cameras. Images \ding{172}, \ding{173} and \ding{174} on the right side show typical emission from each side of microwire when condensate is formed. Dashed lines outline perovskite and substrate boundaries.}
\label{fig:setup}
\end{figure}

To investigate the emission characteristics of perovskite microwires, we built an optical setup that enables simultaneous observation of emission in multiple directions from the microwire.
A schematic representation of the set-up is shown in Figure \ref{fig:setup}.
The microwire is optically excited from the top by Optical Parametric Amplifier (OPA) pumped by pulsed picosecond laser.
We tuned the OPA to emit 1.5 ps pulses with a central wavelength of 435 nm, far above the excitonic resonance in CsPbBr$_3$ \cite{Kędziora-Nature}.
Pump pulses were focused on the perovskite microwire by 50x objective with NA 0.22, resulting in a pump spot size of approximately 8 $\mu$m (defined at the $1/e^2$ intensity level).
Emission from the microwire was simultaneously collected from multiple directions using separate cameras: camera \ding{172} collected emission normal to the substrate (top), camera \ding{173} collected emission along the long axis of the microwire (side), camera \ding{174} collected emission from the microwire end.
A 50/50 beam splitter can be placed in front of any camera to redirect part of the emission into a spectrometer.
Each detection path was equipped with long-pass spectral filter with edge at 473~nm in order to suppress the scattered laser pump.
On the right side of Figure \ref{fig:setup} we show representative emission images acquired with cameras when a polariton condensate is formed within the microwire.
Those images correspond to detection via camera \ding{172} (top), camera \ding{173} (side), and camera \ding{174} (end), respectively.
In each image, we have outlined with dashed lines the boundaries of the perovskite microwire and substrate.

\section{Condensate formation and emission characteristic}

\begin{figure}[htbp]
\includegraphics[width=0.99\linewidth]{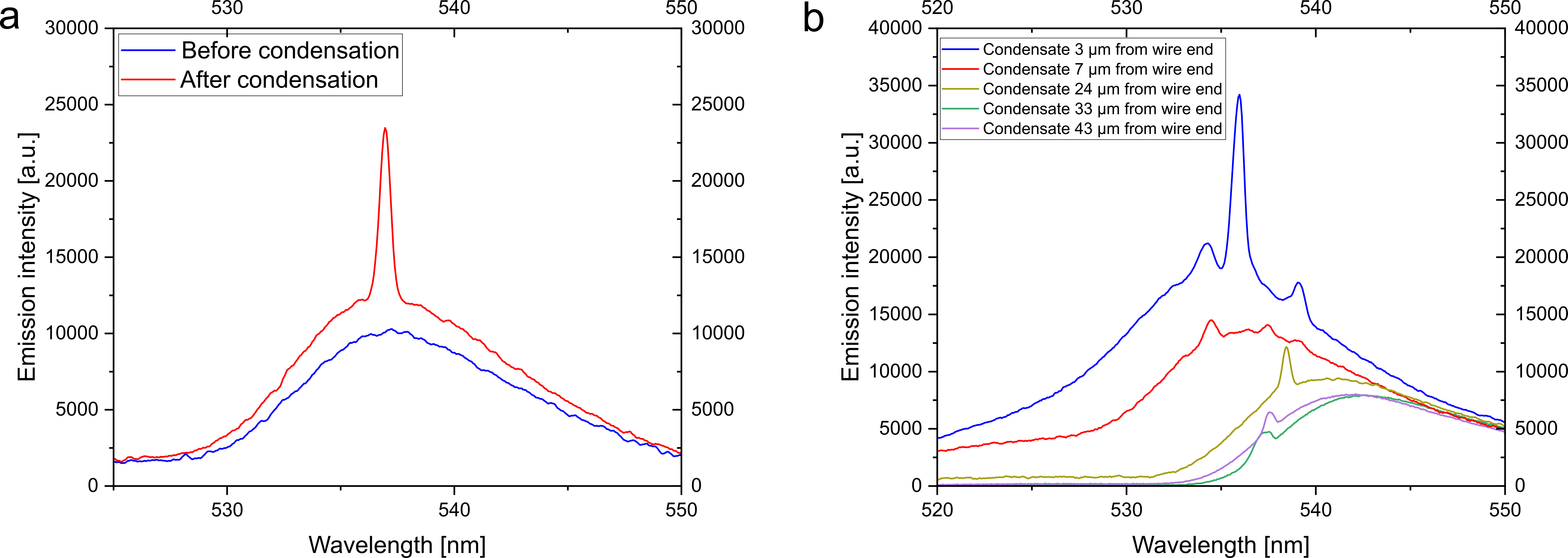}
\caption{Figure \ref{fig:condensat+SS_odl}a change of emission spectrum of perovskite microwire captured from microwire end before and after condensation threshold. Figure \ref{fig:condensat+SS_odl}b change of emission spectrum of perovskite microwire captured from microwire end as the condensate spot is moved further from microwire end.}
\label{fig:condensat+SS_odl}
\end{figure}

Under non-resonant optical excitation at 435~nm the perovskite microwire exhibits spontaneous emission due to the recombination of free excitons in the spectral range of approximately 500 - 560 nm, as illustrated in Figure \ref{fig:condensat+SS_odl}a.
The emission increases linearly with pump pulse energy until reaching the threshold when the exciton-polariton condensate begins to form.
The strong coupling between excitons and photons arises from the natural formation of a Fabry–Pérot resonator across the microwire, enabled by the high refractive index of the perovskite.
Excitons couple to various transverse standing-wave modes, leading to the formation of exciton-polaritons.
The lower polariton branch is energetically shifted from the excitonic resonance due to Rabi splitting, estimated to be 125~meV \cite{Kędziora-Nature}.
At the condensation threshold, polaritons begin to macroscopically occupy a specific optical mode in which gain and loss are balanced, leading to a lasing-like emission from the spontaneously formed polariton condensate.
The condensate exhibits a spectrally narrow emission in green range, and its emission intensity increases nonlinearly with pump pulse energy significantly more rapidly than spontaneous emission.
This nonlinear growth in emission intensity is a hallmark of the phase transition into the condensate state \cite{Fieramosca2019,Su2020-nonlinearroomtemp,Kędziora-Nature}.

\begin{figure}[htbp]
\includegraphics[width=0.99\linewidth]{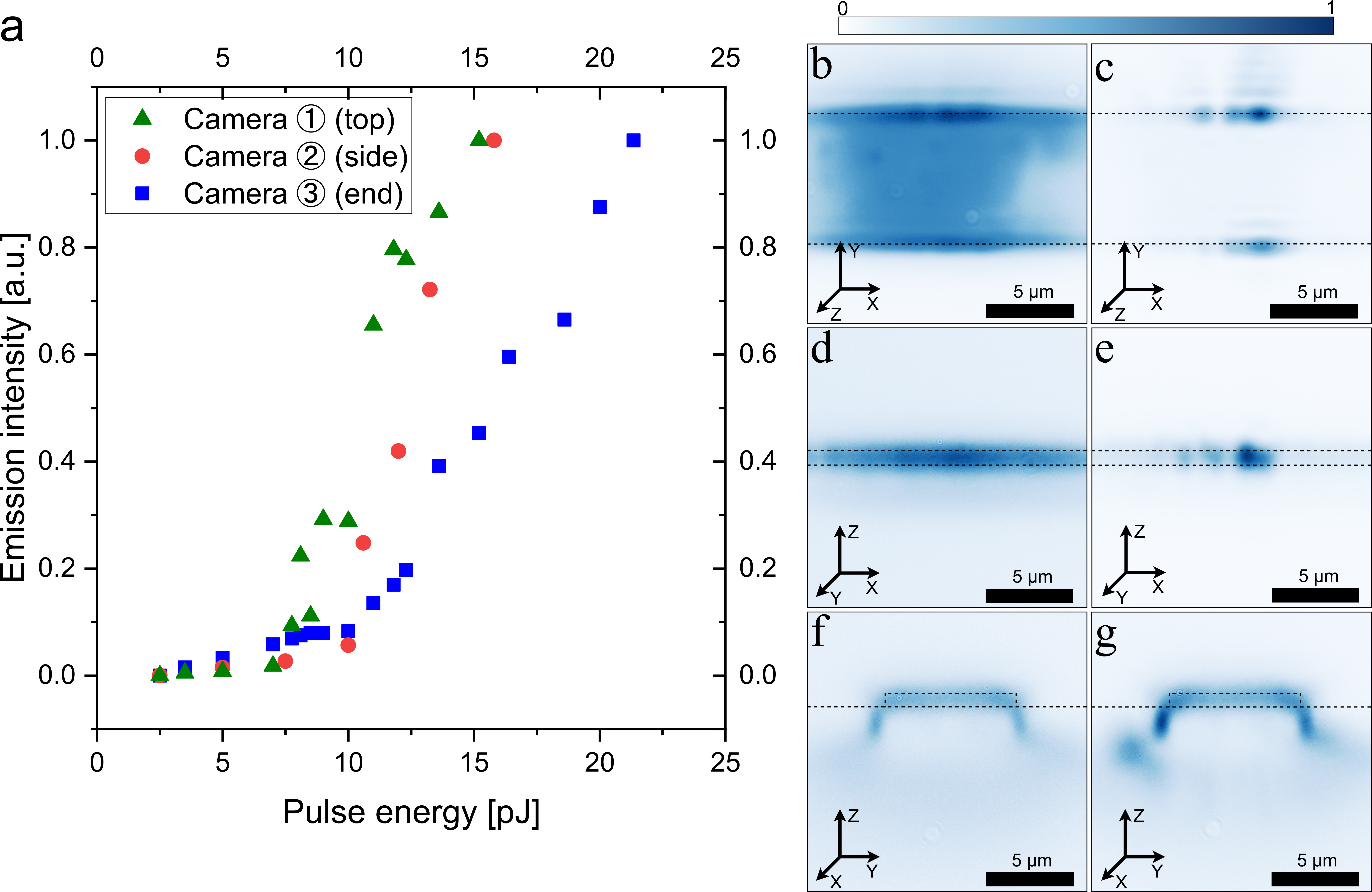}
\caption{Relation between pump pulse energy and perovskite microwire emission intensity measured from three directions. b, d, and f show normalized camera images of microwire emission before condensation for respectively top, side, and end of the microwire. c, e, and g show normalized camera images of microwire emission after condensation for respectively top, side, and end of the microwire.}
\label{fig:progi-all}
\end{figure}

Figure \ref{fig:progi-all}a shows typical emission intensities measured at the condensate wavelength as a function of the excitation pulse energy recorded from the top, side, and end of microwire.
In all three detection configurations, the plots reveal a clear nonlinear threshold behaviour, indicative of condensate formation.

Figures \ref{fig:progi-all}b-g show camera captured intensities before and after condensate formation in the perovskite microwire.
In the top and side views (Figures \ref{fig:progi-all}b-e), the emergence of the condensate leads to a pronounced increase in emission intensity observed at the edges of the microwire, which dominates over the background spontaneous emission.
The condensate formation is far less visible when looking at microwire end as depicted in Figures \ref{fig:progi-all}f-g.
Although a nonlinear increase in emission intensity is still detectable, enabling identification of the condensation threshold, the signal appears weaker compared to top or side detection.

Surprisingly, a strong nonlinear emission is observed in the glass substrate (below dashed line on Figures \ref{fig:progi-all}f-g).
This effect is likely caused by partial leakage of the electromagnetic field from the polariton modes into the substrate, where the evanescent tail of the confined field leads to scattering processes in the glass, enhanced additionally by the imperfections at the microwire–substrate interface.
To reduce the leakage into the substrate, a thin low-refractive-index buffer layer can be introduced between the microwire and the glass, enhancing vertical mode confinement. 
Nevertheless, the primary interest of this study lies in the transport properties of the condensate within the perovskite itself, along the wire.

Figure \ref{fig:condensat+SS_odl} presents the emission spectrum collected from the microwire end. The spectrometer slit was aligned to select only the emission from the microwire itself and block light propagating through the glass substrate.
This spectral analysis offers a clearer view of the condensate emission originating within the microwire even when observed from the end facet.
However, it also reveals that the intensity of the condensate emission (narrow line in Figure \ref{fig:condensat+SS_odl}a) is significantly lower than that of spontaneous emission (broad emission in Figure \ref{fig:condensat+SS_odl}a) under this detection geometry.
As a result, the condensate is less visible when looking at microwire end with camera \ding{174}.
Our cameras integrate over all wavelengths and are therefore dominated by the broader spontaneous emission "background".

The experimental set-up allows for precise translation of the substrate, allowing the excitation spot to be positioned at various locations along the length of the microwire.
This makes it possible to generate the condensate at increasing distances from the microwire end and measure how the emission collected from the end of the microwire evolves as a function of the distance between the condensation spot (pump spot) and microwire end.

Figure \ref{fig:condensat+SS_odl}b illustrates the emission spectra collected from the microwire end as the condensation site position is progressively shifted further away from the facet.
The data reveal a pronounced decrease in emission intensity below 540~nm with increasing distance, indicating that optical absorption in the perovskite material becomes increasingly significant during propagation.
This absorption affects both the broad, incoherent spontaneous emission and the narrow, coherent condensate signal, ultimately limiting the transport efficiency of the guided modes.
Notably, this effect is evident in the attenuation of spectral features at specific wavelengths (e.g., 534~nm and 536~nm) over short distances, while only the longer-wavelength component at 537~nm persists over larger distances and reaches the end of the microwire.

These observations highlight a critical challenge for waveguide design using perovskite materials, where efficient optical transport can only be achieved if the condensate emission lies spectrally outside the material's intrinsic absorption band.
This can be addressed by tailoring the exciton-photon detuning, by chemical composition \cite{Li2021, Kuykendall2007} or waveguide geometry \cite{Michalsky_2018} to shift the condensate emission toward longer wavelengths with lower absorption, while maintaining strong light–matter coupling.

Importantly, despite these losses, condensate emission from the microwire end remains detectable even at a distance of several tens of micrometers from the condensate generation spot.
This demonstrates that the polariton condensate can propagate along the microwire, which is a key requirement for potential applications in integrated photonic and neuromorphic systems.

\section{Effect of structural defects on polariton condensate propagation}

\begin{figure}[htbp]
\includegraphics[width=0.99\linewidth]{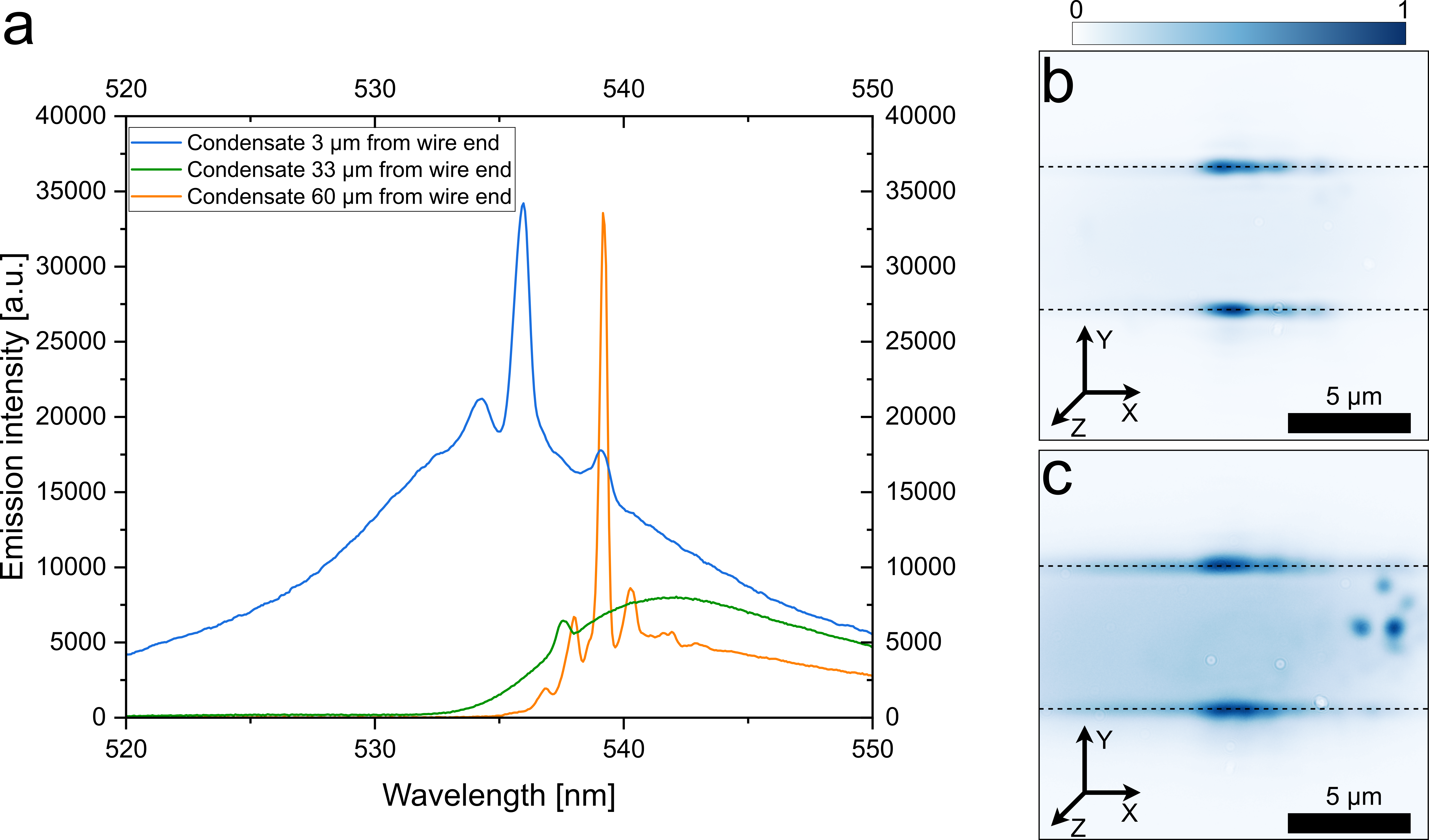}
\caption{Change of emission spectrum of perovskite microwire captured from the short side as the condensate site is shifted away from the end of microwire and closer to defects. The defects are placed 53 µm away from the wire end. a and b show normalized top-side camera emission images at distances of 33 µm and 60 µm away from microwire end, which is to the right of the image. 
}
\label{fig:defekt}
\end{figure}

We observed that the emission intensity of the polariton condensate from microwire end can be significantly enhanced when a structural defect is present near the location of condensate formation. 
This effect is illustrated in Figure \ref{fig:defekt}.

Figure \ref{fig:defekt}a shows the emission spectra collected from the end of the microwire when the polariton condensate is generated at three different distances from the microwire end: 3 µm, 33 µm and 60 µm.
At 33 µm, the polariton condensate emission is nearly undetectable.
However, when the condensate is formed at 60 µm - much greater distance - the polariton condensate emission from the microwire end shows a substantial increase in intensity, despite a continued decrease in the spontaneous emission component. 

This anomalous enhancement is correlated with the presence of a visible defect in the microwire near the polariton condensate region, which appears to play a critical role in modifying the condensate behavior.
Figures \ref{fig:defekt}b and \ref{fig:defekt}c show top-side emission images of condensate formed at 33 µm and 60 µm from the microwire end, respectively.
While there are no noticeable defects visible in Figure \ref{fig:defekt}b, they are clearly visible in Figure \ref{fig:defekt}c.

A plausible explanation is that these defects act as localized optical mode or scattering centers that modify the condensate’s modal structure.
In defect-free regions, polariton condensates typically form in standing-wave modes confined along the transverse (Y) axis of the microwire, between its sidewalls.
The introduction of a defect, however, may induce modes with wavevectors along the longitudinal (X) axis of the microwire, effectively 
scattering part of the condensate into modes aligned along the waveguide direction.
This alignment promotes more efficient forward propagation of the condensate and enhances emission from the microwire end \cite{nano14211691,10.1063/5.0159665,PhysRevB.95.201302,Lingstädt2021}. 

\section{Conclusions}


We have demonstrated exciton-polariton condensation and directional propagation in a CsPbBr$_{3}$ perovskite microwire.
The condensation threshold is characterized by a nonlinear increase in the occupation of lower polariton guided modes, where gain compensates for the losses. 
Our custom multi-angle detection setup-featuring simultaneous top, side, and end-view monitoring enables comprehensive observation of the condensate behavior.
While the condensate emission is most observable in the top and side views, through the scattering of the condensate at the crystal edges, spectral analysis confirms that directional emission from the microwire end persists even when the condensate is formed tens of micrometers away.
Although propagation is limited by absorption losses, condensate emission remains detectable across substantial distances, highlighting the intrinsic waveguiding capability of the microwire.
We also find that structural defects near the condensate region can enhance directional emission by promoting the formation of modes with stronger longitudinal wavevector components.
These findings suggest that condensate transport efficiency could be further improved through the precise engineering of microwire geometry or controlled defect placement.
Our findings show the natural limits of polariton transport, but open promising avenues for perovskite-based integrated photonic and neuromorphic platforms.

\begin{backmatter}
\bmsection{Funding}

\bmsection{Acknowledgment} 
 This work was supported by the National Science Center, Poland, under the following projects: 2020/37/B/ST3/01657, 2022/47/B/ST3/02411,  2021/43/B/ST3/00752, 2023/49/B/ST3/00739 and financed by the European Union EIC Pathfinder Open project “Polariton Neuromorphic Accelerator” (PolArt, Id: 101130304). M.E. and A.Sz. acknowledge the support by the statutory funds of the Łukasiewicz Research Network – Institute of Microelectronics and Photonics.

\bmsection{Disclosures} The authors declare no conflicts of interest.

\bmsection{Data availability} Data underlying the results presented in this paper are not publicly available at this time but may be obtained from the authors upon reasonable request.

\end{backmatter}

\bibliography{bibliography}

\end{document}